\documentclass[aip,10pt,longbibliography,jcp,twocolumn,showpacs,amsmath,amssymb,floatfix]{revtex4-1}
\usepackage{amsmath}
\usepackage{amssymb}
\usepackage{soul}  
\usepackage{url}
\usepackage{hyperref}
\usepackage{subfigure,amsmath,verbatim,moreverb}
\usepackage{multirow}
\usepackage{hhline}
\usepackage{graphicx}  
\usepackage{dcolumn}   
\usepackage{bm}        
\usepackage{amssymb}   
\usepackage{framed}
\usepackage{amsmath}

\begin{document}

\thispagestyle{empty}

\title{Semilocal Exchange Functionals With Improved Performances: The Modified Enhancement Factor For Two Dimensional 
Quantum Systems}

\author{Subrata Jana, Abhilash Patra and Prasanjit Samal}
\affiliation{School of Physical Sciences, National Institute of Science Education and Research,
Bhubaneswar 752050, Homi Bhava National Institute, INDIA}
\date{\today}

\begin{abstract}
Semilocal exchange-correlation functionals are the most accurate, realistic and widely used ones to describe 
the complex many-electron effects of two-dimensional quantum systems. Beyond local density approximation, the 
generalized gradient approximations(GGAs) are designed using reduced density gradient as main ingredient. An 
enhancement factor is constructed using the inhomogeneity parameter of GGAs by taking care of the low and high 
density behaviors of it. Thus, the exchange energy functional proposed by making use of the aforementioned 
enhancement factor, significantly reduces the error compare to the previously proposed gradient approximations. 
Another enhancement factor and corresponding energy functional is also constructed using the inhomogeneity 
parameter originally introduced by Becke [J. Chem. Phys. 109, 2092 (1998)]. Comprehensive testing and performance 
of both the functionals are demonstrated with respect to the exact exchange formalism by considering two-dimensional 
parabolically confined quantum dots with varying particle number and confinement strength as a test case.     
\end{abstract}

\maketitle

\section{\label{sec:intro}Introduction}
In the arena of low-dimensional research, the Hohenberg-Kohn-Sham \cite{hk64,ks65} variant of density-functional 
theory(DFT) is now the most widely applied formalism for electronic-structure calculations. The success of DFT 
is due to the development of several accurate approximations for the exchange and correlation(XC) functionals 
\cite{b83,jp85,pw86,b88,br89,b3pw91,pbe96,kos,vsxc98,hcth,tsuneda,tpss,mO6l,revtpss,tbmbj,scan15,tm16}. The 
applications of DFT are vastly extensive because of the construction of very accurate semilocal density functionals 
\cite{b83,jp85,pw86,b88,br89,b3pw91,pbe96,kos,vsxc98,hcth,tsuneda,tpss,mO6l,revtpss,tbmbj,scan15,tm16}. Despite of 
its grand success, the three dimensional(3D) XC-functionals in principle cannot be extended directly to low-dimensional 
systems due to various limitations \cite{klnlhm}. The development of new functionals in two dimensions(2D), is an 
active area of research with promisingly new perspectives. However, present day studies involving low-dimensional 
systems \cite{kat,rm} e.g. carbenoid, graphene related materials, silicon nanowire based bio-sensors and particularly 
semiconductor layers and surfaces, quantum Hall devices and various types of quantum dots, have keenly attracted the 
attention of researchers and gained momentum. In this regard, many-body effects in low dimensions need to be addressed 
properly for its greater impact in solid-state and materials research. But due to the aforementioned time lag between 
the inception of the 3D and 2D XC functionals, the latter has not been so successful. It is only during the last 
decade or so, increased attention is being paid in developing 2D XC functionals. 

The starting point of 2D XC density functional is obviously the local density approximation(LDA) \cite{rk}. The 
2D-LDA for exchange combined with the 2D correlation \cite{tc,amgb} leads to intriguing results and establishes 
its superiority over quantum Monte Carlo simulations \cite{hser}. Subsequent attempts have also been made to reduce 
the errors of 2D-LDA \cite{prhg,prvm,prg,pr1,prp,rp,sr,pr2,rpvm,prlvm,vrmp}. So generalized gradient approximations
(2D-GGA) \cite{prhg,prvm,prg,pr1,prp,rp,sr,pr2,rpvm,prlvm,vrmp} were the next effective attempts in that direction. 
The 2D-GGA \cite{prvm} reduces the mean percentage error compare to 2D-LDA. The 2D-GGAs have been constructed by 
extending Becke's \cite{b83} proposal to the low dimensional regime. As a matter of which, several reliable and 
accurate semilocal functionals \cite{prhg,prg,pr1,prp,rp,sr,pr2,rpvm,prlvm,vrmp} have been constructed. However, 
non of the above functional have satisfactorily described systems both at the low as well as high density limit.

In DFT, the degree of inhomogeneity associated with the system is included in the construction of XC functionals
through the reduced density gradient($s$), which are the main ingredient of GGA functional. In case of slowly 
varying density, reduced density gradient approaches to zero. Thus, there are two mainstream approaches for
constructing the exchange energy functionals: \cite{b83,b88,br89,tbmbj} and \cite{pw86,pbe96,tpss,revtpss,scan15}. 
The functionals proposed by Becke, \cite{b88,br89} contain exchange hole potential and using it one in principle 
can construct the corresponding energy functionals. But, in these cases the potentials are not the functional 
derivative of exchange energies. Whereas, in case of functionals proposed by Perdew \cite{pw86,pbe96,tpss,revtpss,
scan15} et al are based on the enhancement factors along with LDA for XC. So in contrast, to Becke's approach, 
the XC potentials are nothing but the functional derivative of the corresponding functionals. On applying spin 
density scaling, one can easily construct its spin polarization version. Unlike GGA, the meta-GGA exchange energy 
functionals use the non-interacting positive definite KS kinetic energy density ($\tau$) and '$s$' as its ingredients. 
Thus '$s$' together with '$\tau$' forms the higher order rung of XC functionals. In stead of '$\tau$', Becke proposed 
that a new inhomogeneity parameter \cite{b98} can be used to construct the XC-functionals. The present work aims at 
constructing reliable and most appropriate enhancement factors using '$s$' and Becke's inhomogeneity parameter 
for 2D quantum systems. So using the above enhancement factors we have proposed two semilocal exchange functionals.

This work is organized as follows: In the next section, we will briefly discuss exchange hole and its connection with 
exchange energy. This will be used in the following section to construct the low and high density limit of enhancement 
factor. Then, we will propose a form for the enhancement factor through extrapolation between the low and high density 
limit. To fit and test the performance of the newly constructed functional, it'll be applied to study few electron 
quantum dots. An inhomogeneity parameter based on coordinate transformation is also proposed which can be further used 
to construct series of enhancement factors and functionals. In the appendices, we'll illustrate a scheme for constructing 
potentials for GGA and meta-GGA energy functional used in the present work. 

\section{\label{sec:method}Exchange Hole and Exchange Energy}
The exchange energy is considered as the electrostatic interaction between the electron at $\vec{r}$ with the 
exchange hole at $\vec{r}+\vec{u}$ surrounding the electron. So the spin-unpolarized exchange energy can be 
defined as
\begin{equation}
E_x[\rho]=\frac{1}{2}\int~d^2r\int~d^2u\frac{\rho(\vec{r})\rho_x(\vec{r},\vec{r}+\vec{u})}{u}~.
\label{eq1}
\end{equation} 
The exchange hole appearing in Eq.(\ref{eq1}), is associated with the $1^{st}$ order reduced density matrix and is
given by 
\begin{equation}
\rho_x(\vec{r},\vec{r}+\vec{u})=-\frac{|\varGamma(\vec{r},\vec{r}+\vec{u})|^2}{2\rho(\vec{r})}~,
\label{eq2}
\end{equation}
with $\varGamma(\vec{r},\vec{r}+\vec{u})=2\sum_{i}^{occ}\psi_i^*(\vec{r})\psi_i(\vec{r}+\vec{u})$, where
$\psi_i$ are the occupied KS orbitals. The exchange hole has two important properties: 
(i) the normalization sum rule $\int~\rho_x(\vec{r},\vec{r}+\vec{u})~d^2u = -1$ and (ii) the negativity 
constraint $\rho_x(\vec{r},\vec{r}+\vec{u})\leq0$. In $2D$, the exchange energy, $E_x$ involves the 
cylindrical average of the exchange hole, $\langle\rho_x(\vec{r},\vec{r}+\vec{u})\rangle_{cyl}$ over the 
direction of $\vec{u}$, i.e.
\begin{equation}
\langle\rho_x(\vec{r},\vec{r}+\vec{u})\rangle_{cyl}=\int\frac{d\Omega_u}{2\pi}\rho_x(\vec{r},\vec{r}+\vec{u})~.
\label{eq3}
\end{equation}
Using spin-scaling relation, the exchange energy functional can be easily generalized to any spin polarization, 
i.e.
\begin{equation}
E_x[\rho_\uparrow,\rho_\downarrow]=\frac{1}{2}E_x[2\rho_\uparrow]+\frac{1}{2}E_x[2\rho_\downarrow]~.
\label{eq4}
\end{equation}
The exchange energy functional also satisfies the uniform coordinate scaling property i.e.
\begin{equation}
E_x[\rho_\gamma]=\gamma E_x[\rho],
\label{eq5}
\end{equation}
where $\rho_\gamma=\gamma^2\rho$ is the scaling of the electronic density. Since in terms of the enhancement 
factor, the GGA functional is given by
\begin{equation}
E_x^{GGA}[\rho] = \int~d^2r~A_x\rho(\vec{r})^{3/2}F_x[s]~,
\label{eq6}
\end{equation}
where $A_x=\frac{4(2\pi)^{1/2}}{3\pi}$ and reduced density gradient, $s=\frac{|\vec{\nabla}\rho|}{2(2\pi)^{1/2}
\rho^{3/2}}$ (which is the main ingredient of GGA functional). Thus, the functional, $F_x(s)$ must reduces to 
unity ($1$) when $s=0$, in order to recover the correct exchange energy for uniform density i.e. LDA. Actually, 
there happens to be several ways of constructing enhancement factor $F_x$: (i) it can be constructed by using the 
small and large gradient approximations of $F_x$ and then by employing extrapolation between these two limits, 
(ii) by using properties of exchange potential or exchange energy and (iii) by imposing relevant physical constraints. 
The GGA constructed by R\"{a}s\"{a}nan et. al. \cite{prvm} used the approach (i). Later 2D-B88 \cite{vrmp} formed by 
applying approach (ii). Here, in this case we have constructed two new semilocal exchange functionals by employing 
the approach (i). For doing that, we'll now elaborate on the low and high density limits of the enhancement factor.  

\subsection{\label{sec:cusp}Small Gradient Behavior}
Lets begin with the small gradient expansion of the enhancement factor for exchange energy in 2D. To do this, we have 
revisited the formalism originally proposed by Becke \cite{b83} in 3D and the extension of it to 2D \cite{prvm}. As 
the Taylor series expansion of cylindrical averaged conventional (because no coordinate transformation is involved.) 
exchange hole is
\begin{equation}
\begin{split}
\langle\rho_{x2D}\rangle = -\frac{\rho(\vec{r})}{2}-\frac{1}{4}\Big[\frac{1}{2}\nabla^2\rho(\vec{r})-2\tau
+\frac{1}{4}\frac{|\vec{\nabla}\rho(\vec{r})|^2}{\rho(\vec{r})}\Big]u^2~.
\end{split}
\label{eq7}
\end{equation}
Now, an exchange hole, based on coordinate transformation can also be proposed which is given in Appendix-(\ref{ap1}). 
For small inhomogeneity, one can consider 2D homogeneous electron gas (2D-HEG) as a good reference system. Then, the 
cylindrical averaged uniform exchange hole is given by
\begin{equation}
\langle\rho_{x2D}^{unif}\rangle = \frac{2J_1^2(k_Fu)}{k_F^2u^2}\rho(\vec{r})~,
\label{eq8}
\end{equation}
where $u$ be the separation between pair of electrons and $k_F = (2\pi\rho)^{\frac{1}{2}}$ is the Thomas-Fermi 
wavevector in 2D. So the cylindrical averaged exchange hole can be expressed in terms of the polynomials of $u$,
\begin{equation}
\langle\rho_{x2D}\rangle = \Big[1 + a(\vec{r})u^2 + b(\vec{r})u^4 + ........\Big]\langle\rho_{x2D}^{unif}\rangle~.
\label{eq9}
\end{equation}
Now, truncating the polynomial up to $u^4$ and comparing it with the Taylor series expansion of the cylindrical averaged 
exchange hole i.e. Eq.(\ref{eq7}) leads to
\begin{equation}
a(\vec{r}) = \frac{1}{2\rho}\Big[\frac{1}{3}\nabla^2\rho + \frac{1}{4}\frac{|\vec{\nabla}\rho|^2}{\rho}\Big].
\label{eq10}
\end{equation}
However, by applying normalized sum rule constraint to the cylindrical averaged exchange hole, the coefficient $b$ 
turns out to be
\begin{equation}
b(\vec{r})=-2\pi\frac{I(1)}{I(3)}\rho(\vec{r})a(\vec{r})~,
\label{eq11}
\end{equation}
where $I(m)$ is nothing but
\begin{equation}
I(m) = \int_0^z dyJ_1^2(y)~,
\label{eq12}
\end{equation} 
with $z$ corresponding to the $1^{st}$ zero of the Bessel function and has to be evaluated numerically. For slowly varying 
density, semi-classical approximation of kinetic energy density can be used and upon substituting it in the enhancement 
factor, the same modifies to
\begin{eqnarray}
F_x^{SGL}&=& 1 + \mu^{SGL} s^2~,
\label{eq13}
\end{eqnarray}
where $s = \frac{|\nabla\rho|}{2k_F\rho}$ be the reduced density gradient. Whereas, $\mu=3\pi^{3/2}\kappa^{SGL}$ is the 
small gradient coefficient of the enhancement factor, with $\kappa^{SGL}=4\tilde{\kappa}/6=0.0072452$ and $\tilde{\kappa}$ 
obtained from $\frac{1}{4^{3/2}\sqrt{\pi}}\Big[\frac{I(0)I(3)-I(1)I(2)}{I(3)}\Big]$.

\subsection{\label{sec:asymp}Large Gradient Behavior}
The large gradient behavior of exchange hole as discussed by Becke's \cite{b83} and R\"{a}s\"{a}nan \cite{prvm} which
give rise to 
\begin{equation}
\langle\rho_{x2D}\rangle\approx\Big[\frac{1}{4}\frac{|{\vec{\nabla}}\rho|^2}{\rho}~R^2\Big]e^{-\alpha(\vec{r})^2R^2}~.
\label{eq14}
\end{equation} 
This Gaussian approximation of exchange hole is proposed in order to produce correct short-range behavior and finite 
exchange energy at large density gradient. The parameter $\alpha$, is obtained from the normalization condition of 
exchange hole and is given by
\begin{equation}
\alpha(\vec{r}) = \Big[\frac{\pi G(3)}{2}\frac{|\vec{\nabla}\rho|^2}{\rho}\Big]~,
\label{eq15}
\end{equation}
with
\begin{equation}
G(m)=\int_{0}^{\infty}~dy~y^m~e^{-y^2}~.
\label{eq16}
\end{equation}
On using these large gradient limit results, the new enhancement factor can be obtained as
\begin{eqnarray}
F_x^{LGL}&=& 1 + \mu^{LGL} s^\frac{1}{2}~.
\label{eq17}
\end{eqnarray}
The factor $\mu^{LGL}=(\frac{\pi}{4})^{\frac{1}{4}}\frac{\kappa^{LGL}}{A_x}$ is the large gradient coefficient 
of the enhancement factor, with $A_x=\frac{4(2\pi)^{\frac{1}{2}}}{3\pi}$. The parameter $\kappa^{LGL}$ is obtained 
to be $0.35078$.

\subsection{\label{sec:enhancement}The Modified Enhancement Factor}
It is trivial from the SGL and LGL of the enhancement factor that $F_x$ behaves as $s^2$ and $s^{\frac{1}{2}}$ 
respectively. Now we'll combine these LGL and SGL results to find an analogous and more general expression for 
enhancement factor that interpolates between the two limits. A possible expression is
\begin{equation}
F^{MOD-GGA}_x(s)=1+\mu\frac{s\log(g)}{1+\beta s^{\frac{1}{2}}\log(g)+(1-e^{-c s^2})}
\label{eq18}
\end{equation}
with $g = s + \sqrt{1+s^2}$. This form obeys the large and small gradient behaviors of enhancement factor, though the 
parameters are different from its SGL and LGL value as obtained above. The three parameters $\mu$, $\beta$ and $c$ 
are obtained by using LGL of enhancement factor by considering the physically relevant 2D systems like few electrons 
parabolic quantum dots. As matter of which, the parameters $\mu$, $\beta$ and $c$ are obtained to be $0.84089~\mu^{LGL}$, 
$0.248$ and $0.1$ respectively by employing the exact KLI-OEP result of 2D-quantum dots. By virtue of the above parameters, 
the mean percentage error for the overall test set gets reduced. Next, the new exchange functional constructed by employing 
the above enhancement factor is also applied to same set of parabolically confined quantum dots. As a reference set, we have
performed self-consistent KLI-OEP calculations. Also the KLI-OEP density is used as reference input for the testing the
performance of the newly developed functional. Then, the results are also compared w.r.t. 2D-LDA, 2D-B88  \cite{vrmp} and 
2D-GGA \cite{prvm}. The results obtained with the new functional are given in the Table-(I) which confirms the significant
amount of reduction of error compare to the GGA functionals. Henceforth, we name this functional as modified GGA (MOD-GGA) 
as it is constructed by making modification over the existing GGA functional.

\begin{figure}
\begin{center}
\includegraphics[width=3.4in,height=2.0in,angle=0.0]{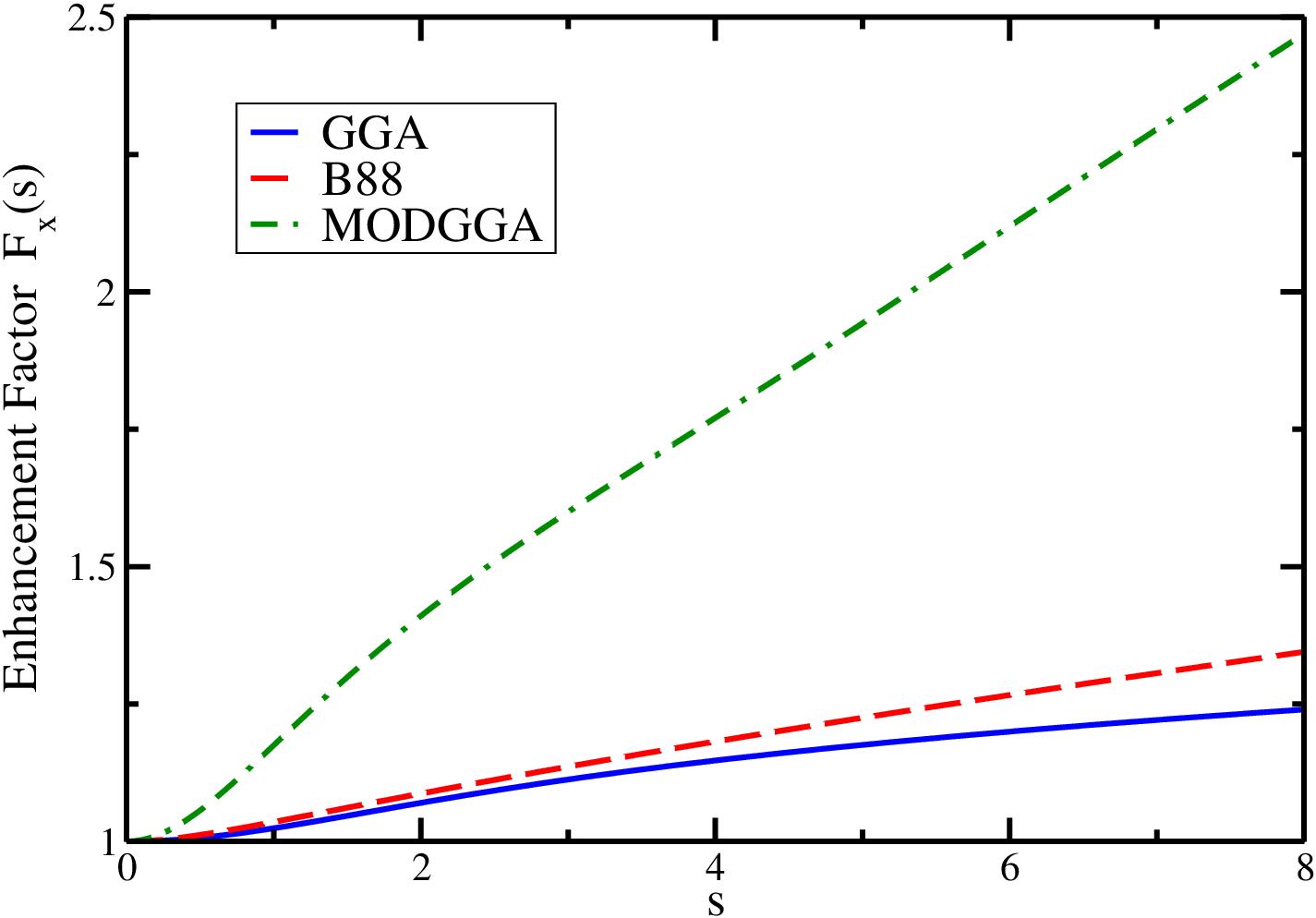} 
\end{center}
\caption{Plotted are the enhancement factors $F_x$ of the MOD-GGA functional and that of 2D-B88 and 2D-GGA for comparison.}
\label{fig1}
\end{figure}

\begin{table}\label{t1}
\caption{Shown are exchange energies (in atomic units) for parabolically confined few-electron quantum dots using 
self-consistent calculation with OCTOPUS code \cite{octopus}. The $1^{st}$ column contains the number of particles. 
The $2^{nd}$ column contains different confinement strength used for fitting the parameters of the new functional.
For the MOD-GGA calculation, the output spin polarized density and kinetic energy density of KLI-OEP is used. The
results of the $MOD-GGA$ functional is presented at the last column. Results for 2D $EXX$, $LDA$, $GGA$ \cite{prvm} 
and $B88$) \cite{vrmp} are also shown for comparison. The last row contains the mean percentage error, $\Delta$.}
\begin{tabular}{c  c  c  c  c  c   c   c}
\hline\hline
N&$\omega$&$-E_x^{EXX}$&$-E_{x}^{LDA}$&$-E_x^{GGA}$&$-E_x^{B88}$ &$-E_x^{MODGGA}$\\ \hline
2& 1/6&0.380&0.337& 0.368&0.364 &0.378  \\
2& 0.25&0.485&0.431 & 0.470&0.464& 0.482 \\
2& 0.50&0.729&0.649 & 0.707&0.699 & 0.723 \\
2& 1.00& 1.083&0.967& 1.051 &1.039 &1.070 \\
2& 1.50&1.358&1.214& 1.319&1.304  & 1.361  \\
2& 2.50&1.797&1.610&1.748&1.728  & 1.756 \\
2& 3.50&2.157&1.934&2.097&2.074  & 2.089 \\
6& $1/1.89^2$&1.735&1.642&1.719&1.775& 1.735\\
6& 0.25&1.618&1.531 &1.603&1.594  &1.619    \\
6& 0.42168&2.229&2.110 &2.206&2.241  &  2.228  \\
6& 0.50&2.470&2.339&2.444&2.431& 2.469  \\
6& 1.00&3.732&3.537&3.690&3.742& 3.727\\
6& 1.50&4.726&4.482&4.672&4.648& 4.716  \\
6& 2.50&6.331&6.008&6.258&6.226&  6.305 \\
6& 3.50&7.651&7.264&7.562&7.525& 7.605  \\
12&0.50&5.431&5.257&5.406&5.387& 5.434   \\
12& 1.00&8.275&8.013& 8.230&8.311& 8.275\\
12& 1.50&10.535&10.206&10.476&10.444& 10.518  \\
12& 2.50&14.204&13.765&14.122&14.080& 14.149  \\
12& 3.50&17.237&16.709&17.136&17.086& 17.129  \\
20&0.50&9.765&9.553&9.746&9.722& 9.780  \\
20& 1.00&14.957&14.638&14.919& 15.029& 14.970\\
20& 1.50&19.108&18.704&19.053&19.188&19.113  \\
20& 2.50&25.875&25.334&25.796&25.973&25.853   \\
20& 3.50&31.491&30.837&31.392&31.603&31.429  \\
30& 1.00&23.979&23.610&23.953&24.091&24.000   \\
30& 1.50&30.707&30.237&30.665&30.836&30.813   \\
30& 2.50&41.718&41.085&41.651&41.878&41.675   \\
30& 3.50&50.882&50.115&50.794&51.068&50.763  \\
42& 1.00&35.513&35.099&35.503&35.671&35.557   \\
42& 1.50&45.659&45.032&45.538&45.747&45.600   \\
42& 2.50&62.051&61.339&62.007&62.286&62.053   \\
42& 3.50&75.814&74.946&75.748&76.085&75.758  \\
56& 1.00&49.710&49.256&49.722&49.919&49.769   \\
56& 1.50&63.869&63.289&63.871&64.117&64.050   \\
56& 2.50&87.164&86.378&87.148&87.479&87.150   \\
56& 3.50&106.639&105.684&106.609&107.010&106.527  \\
72& 1.00&66.708&66.219&66.746&66.972&66.796   \\
72& 1.50&85.814&85.186&85.844&86.129&85.898   \\
72& 2.50&117.312&116.456&117.327&117.712&117.352   \\
72& 3.50&143.696&142.650&143.697&144.163&143.657  \\
90& 1.00&86.631&86.111&86.698&86.954&86.737   \\
90& 1.50&111.558&110.889&111.622&111.946&111.655   \\
90& 2.50&152.723&151.808&152.779&153.217&152.750   \\
90& 3.50&187.262&186.139&187.306&187.838&187.164  \\
110& 1.00&109.595&109.048&109.695&109.981&109.736   \\
110& 1.50&141.255&140.548&141.357&141.720  &141.395   \\
110& 2.50&193.617&192.647&193.715&194.210   &193.705   \\
110& 3.50&237.612&236.420&237.706&238.306&237.589  \\
\hline\hline
$\Delta$& & &5.36 &0.71 &2.60 &0.29 \\
\hline\hline
\end{tabular}
\end{table}
\section{\label{sec:ing}Enhancement Factor From Becke's Inhomogeneity Parameter}
Becke showed that \cite{b98} the coefficient of $u^2$ of the Taylor series expansion of exchange hole in Eq.(\ref{eq7}) 
is a "self-interaction" free term i.e., the interaction between the electron and the hole surrounding it at each reference 
point is zero for one electron. For one electron, the kinetic energy term present within the square bracket of Eq.(\ref{eq7}) 
exactly cancels with the gradient term. Thus an important inhomogeneity parameter can be given along with reduced density 
gradient, $Q_B$, as
\begin{equation}
Q_B = \frac{1}{\tau_0}\Big[\tau_0-\tau+\frac{1}{8}\frac{|\vec{\nabla}\rho|^2}{\rho}+\frac{1}{4}\nabla^2\rho\Big]
\label{eq19}
\end{equation}
with,
\begin{equation}
\tau_0=\frac{1}{4}k_F^2\rho~.
\label{eq20}
\end{equation}

\begin{figure}
\begin{center}
\includegraphics[width=3.4in,height=2.0in,angle=0.0]{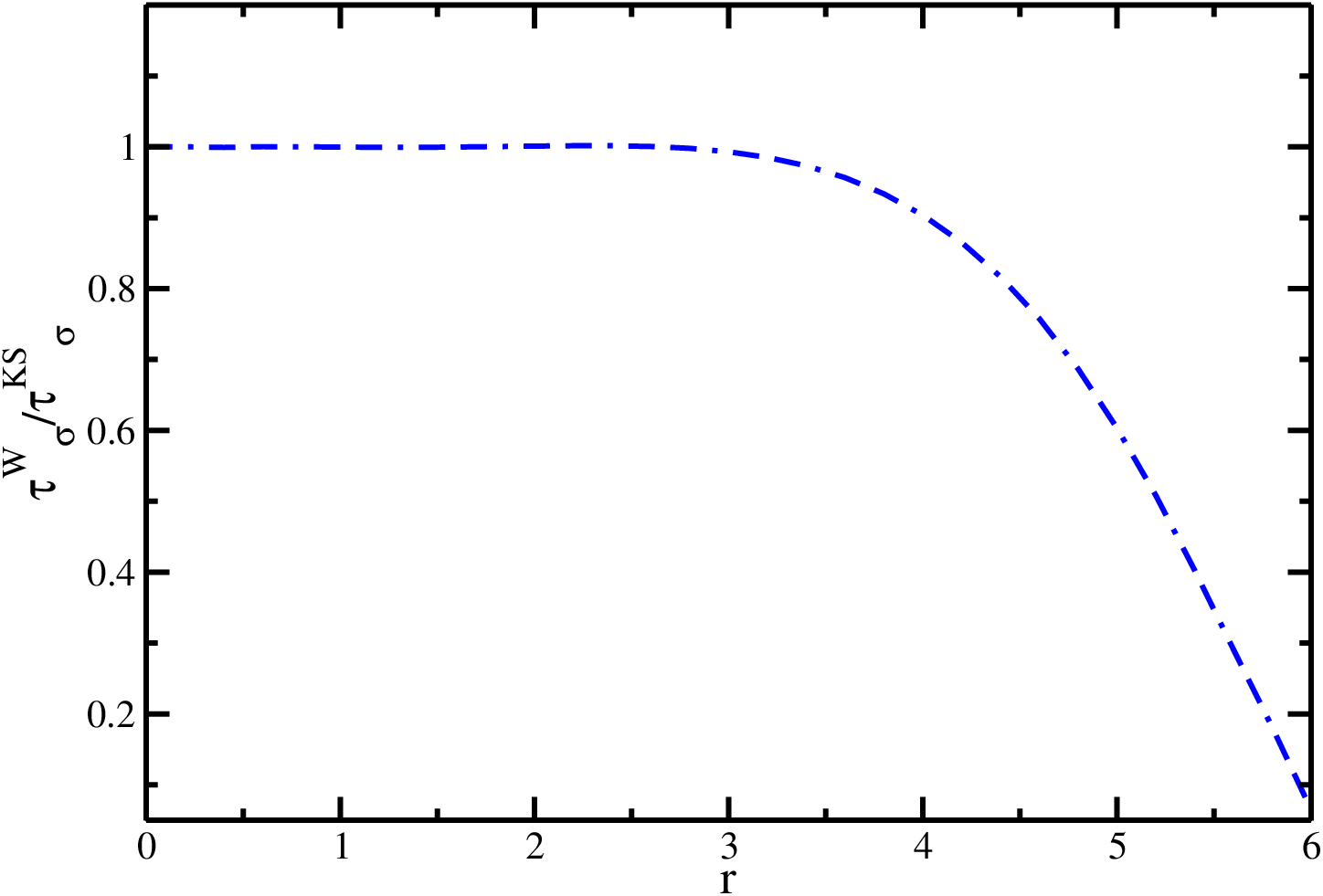} 
\end{center}
\caption{Ratio of spin polarized von Weizs\"{a}cker (WV) kinetic energy density ($\tau^W_{\sigma}=\frac{|\vec{\nabla}
\rho_{\sigma}|^2}{4\rho_{\sigma}}$) to the exact kinetic energy density ($\tau^{KS}_{\sigma}$) for two electrons 
confined in a parabolic quantum dot of confinement strength $\omega=1$. This figure shows that near origin 
$\tau^{KS}_{\sigma}\to\tau^W_{\sigma}$.}
\label{fig3}
\end{figure}

This inhomogeneity parameter can be used to define the diffuse or compact exchange hole surrounding an electron. If 
the exact quadratic term i.e. the term containing inhomogeneity parameter is larger than the homogeneous counterpart 
then it represents more compact hole if not then diffuse exchange hole. The inhomogeneity parameter is zero for uniform 
density. Near the origin, $\tau^{KS}\approx\tau^W$ (see figure-(\ref{fig3})) and therefore it depends only on the Laplacian 
of density. Also in the exponential tail region as the KS KE density equals to the VW correction, due to its one electron 
like character, it depends only on the Laplacian of density. For two dimensional quantum system , near $r\to 0$ the 
Laplacian of density is finite but in exponential tail it tends to $\infty$. It is also invariant under uniform coordinate 
scaling, i.e.,
\begin{equation}
Q_B[\rho_\lambda;\vec{r}] = Q_B[\rho;\lambda\vec{r}]~.
\label{eq21}
\end{equation}
In the intermediate region, it becomes positive.

\subsection{\label{sec:modeling}Modeling The Enhancement Factor}
An enhancement factor cab be designed using the property of the $Q_B$.
For slowly varying density, $Q_B$ is small. Thus, to recover the gradient expansion of the enhancement factor, one
may Taylor expand $F_x(Q_B)$ as a power series of $Q_B$ about $Q_B=0$:
\begin{equation}
F_x(Q_B)=F_x(Q_B)|_{Q_B=0} + F'_x(Q_B)|_{Q_B=0}~Q_B+.....
\label{eq23}
\end{equation}
where
\begin{equation}
F'_x(0) = \frac{dF_x}{dQ_B}\Big|_{Q_B=0}~.
\label{eq24}
\end{equation}
As for slowly varying density, the gradient expansion of kinetic energy density is given by,
\begin{equation}
\tau^{GEA} = \tau_0 + \frac{1}{6}\nabla^2\rho~.
\label{eq25}
\end{equation}
Therefore,
\begin{eqnarray}
Q_B&=&\frac{1}{\tau_0}\Big[\tau_0-\tau_0+\frac{1}{8}\frac{|\vec{\nabla}\rho|^2}{\rho}+\frac{1}{4}\nabla^2\rho\Big]
\nonumber\\
&=&\frac{1}{\tau_0}\Big[\frac{1}{8}\frac{|\vec{\nabla}\rho|^2}{\rho}+\frac{1}{12}\nabla^2\rho\Big]\nonumber\\
&=&\frac{2}{3}\Big[3p+2q\Big],
\label{eq26}
\end{eqnarray}
where two dimensionless parameter $p$ and $q$ are defined as reduced density gradient and reduced Laplacian density 
gradient given by
\begin{equation}
p = \frac{|\vec{\nabla}\rho|^2}{(2k_F\rho)^2};~~~~ q = \frac{\nabla^2\rho}{4k_F^2\rho}~.
\label{eq27}
\end{equation}
Thus,
\begin{equation}
F_x[p,q] = 1+\frac{2}{3}F'_x(0)(3p+2q)
\label{eq28}
\end{equation}
with the corresponding exchange energy functional given by
\begin{equation}
E_x[\rho]=\int~d^2r\epsilon_x^{LDA}F_x[p,q]~.
\label{eq29}
\end{equation}
Now to eliminate the Laplacian we use integrating by parts, so that
\begin{equation}
\int~d^2r\rho(\vec{r})\epsilon_x^{LDA}q = \frac{1}{2}\int~d^2r\rho(\vec{r})\epsilon_x^{LDA}p~.
\label{eq30}
\end{equation}
As a matter of which, the enhancement factor becomes
\begin{eqnarray}
F_x[p] &=& 1+\frac{2}{3}F_x'(0)(3p+p)\nonumber\\
       &=& 1+\frac{8}{3} p F_x'(0)
\label{eq31}
\end{eqnarray}
For slowly varying density limit. From Eq.(\ref{eq13}),
\begin{equation}
F_x[p] = 1+\mu^{SGL}p~.
\label{eq32}
\end{equation}
Comparing Eq.(\ref{eq31}) and Eq.(\ref{eq32}), we have obtained $F_x'(0)=\frac{3}{8}\mu^{SGL}$. So the simplest conceivable 
enhancement factor is,
\begin{equation}
F_x[Q_B] = 1 + \frac{\alpha Q_B}{\sqrt{1 + (\gamma Q_B)^2}}
\label{eq33}
\end{equation}
with $\alpha = \frac{3}{8}\mu^{SGL}$ and $\gamma$ value need to be chosen so as to reduce the mean percentage error for the 
overall test set.

\subsection{\label{sec:performance}Performance Of The Functional}
To test the accuracy and efficiency of the newly constructed functional described above, we have applied it to the few 
electron parabolic quantum dot. From the test set we have chosen the value of $\gamma$ to be $0.0001$. This functional 
has been tested along with GGA and meta-GGA type functionals such as Becke-Roussel \cite{prg}. The results are shown in 
Table - II, where the new functional is denoted as $MGGA$.

\begin{table}[!htb]
\caption{The table caption is same as Table-I except that comparison of results are done w.r.t. 2D $EXX$ and $BR$
\cite{prhg}.}
\begin{tabular}{c  c  c  c  c c }
\hline\hline
N&$\omega$&$-E_x^{EXX}$ &$-E_x^{BR}$ &$E_x^{MGGA}$\\ \hline
2& 1/6&0.380&0.375 & 0.381  \\
2& 0.25&0.485&0.480& 0.485  \\
2& 0.50&0.729&0.722&  0.724 \\
2& 1.00& 1.083&1.080&1.069 \\
2& 1.50&1.358&1.354 &1.334   \\
2& 2.50&1.797&1.794 & 1.749  \\
2& 3.50&2.157&2.020 & 2.078  \\
6& $1/1.89^2$&1.735&1.775& 1.756 \\
6& 0.25&1.618&1.655 &  1.639  \\
6& 0.42168&2.229&2.281 & 2.251   \\
6& 0.50&2.470&2.529&  2.494 \\
6& 1.00&3.732&3.824& 3.755\\
6& 1.50&4.726&4.845&  4.747 \\
6& 2.50&6.331&6.492& 6.343  \\
6& 3.50&7.651&7.846&  7.650 \\
12&0.50&5.431&5.728&  5.457  \\
12& 1.00&8.275&8.572& 8.293\\
12& 1.50&10.535&10.915& 10.540  \\
12& 2.50&14.204&14.716& 14.168  \\
12& 3.50&17.237&17.858& 17.148  \\
20&0.50&9.765&10.167&   9.819 \\
20& 1.00&14.957&15.573& 15.013\\
20& 1.50&19.108&19.892& 19.159  \\
20& 2.50&25.875&26.935& 25.905  \\
20& 3.50&31.491&32.777&3 31.483  \\
\hline
$\Delta$& & & 2.58& 0.75\\
\hline\hline
\end{tabular}
\end{table}

\begin{figure*}
\centering
  \begin{tabular}{@{}cccc@{}}
    \includegraphics[width=3.4in,height=2.0in,angle=0.0]{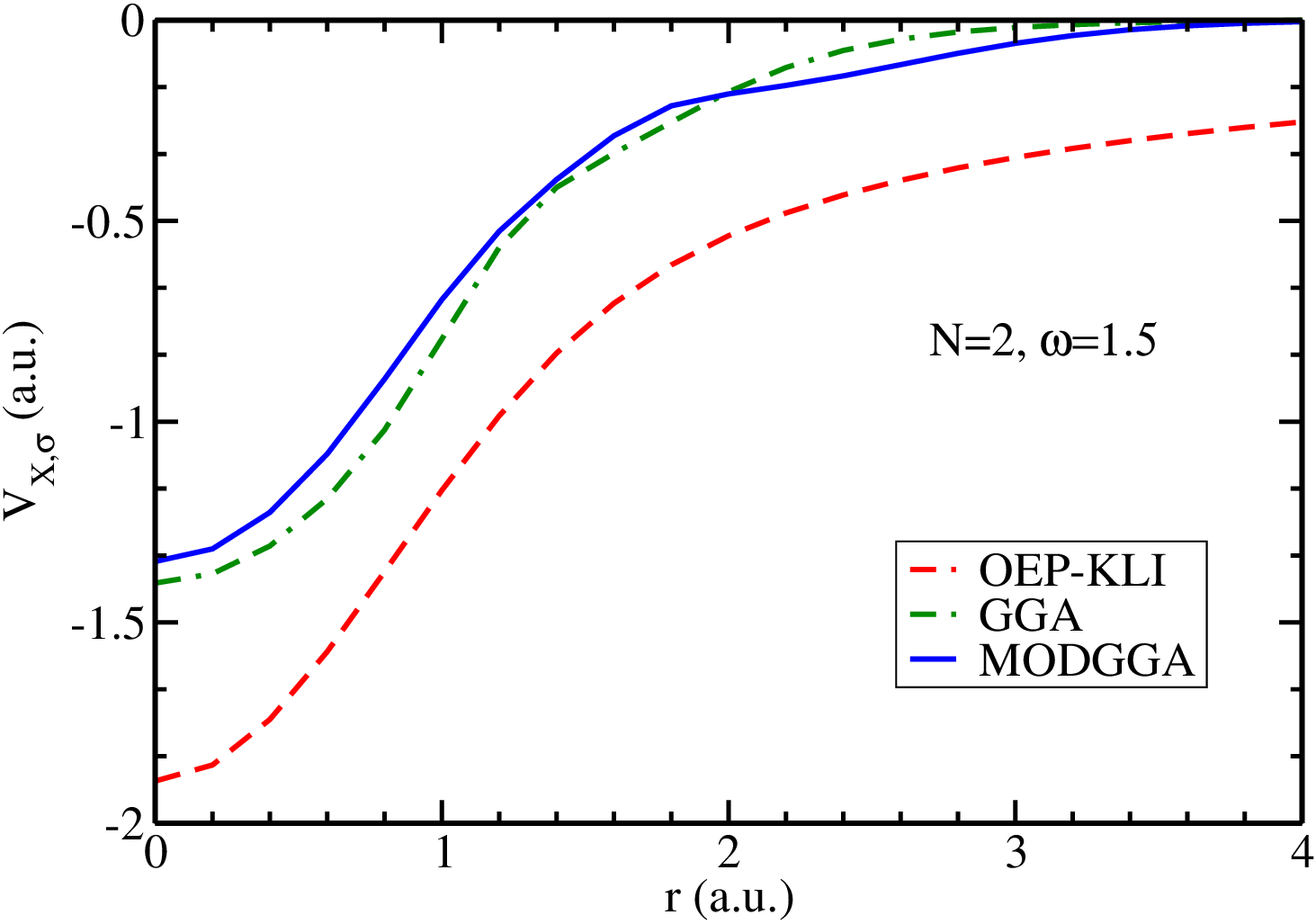} &
    \includegraphics[width=3.4in,height=2.0in,angle=0.0]{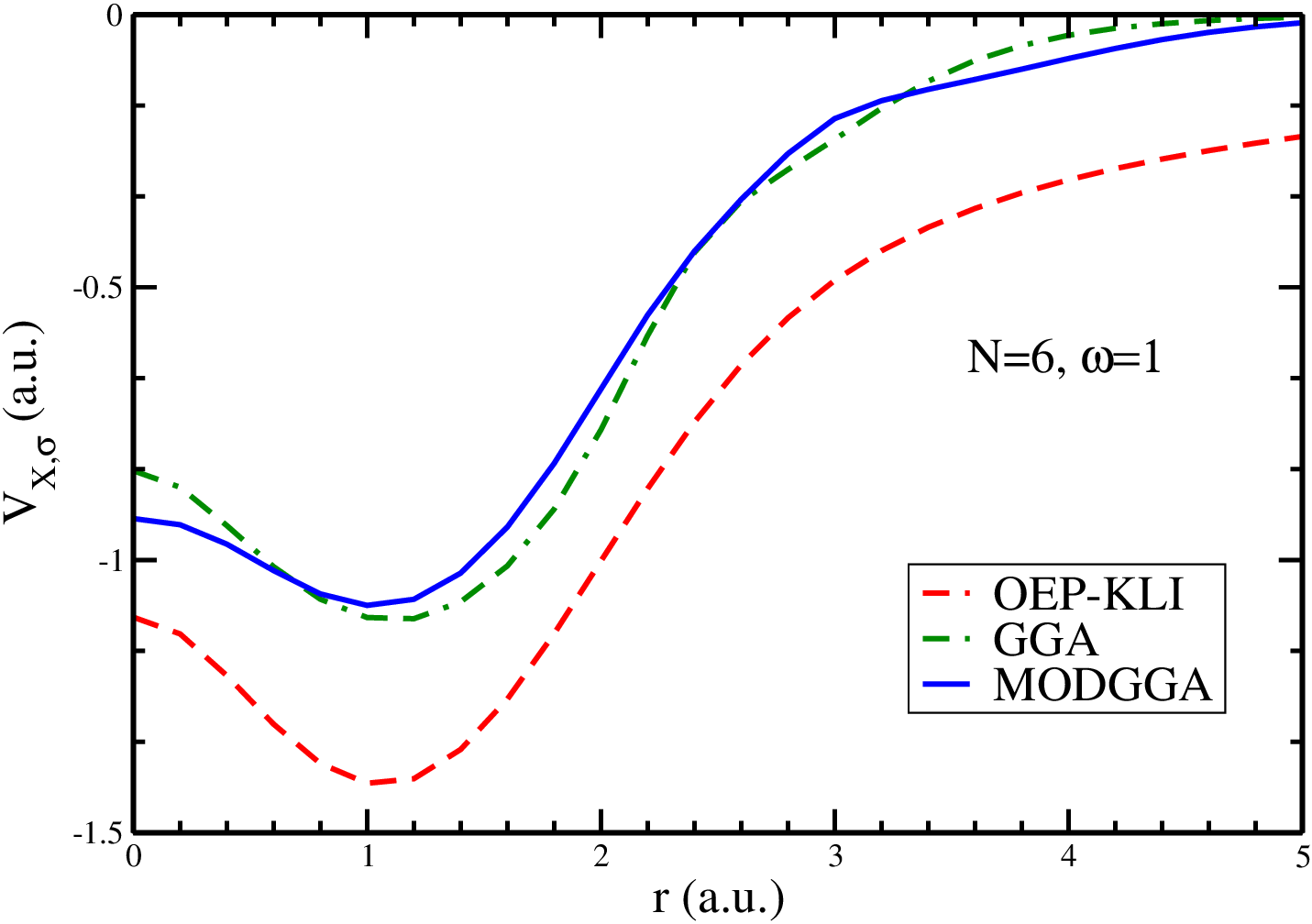} \\
    \includegraphics[width=3.4in,height=2.0in,angle=0.0]{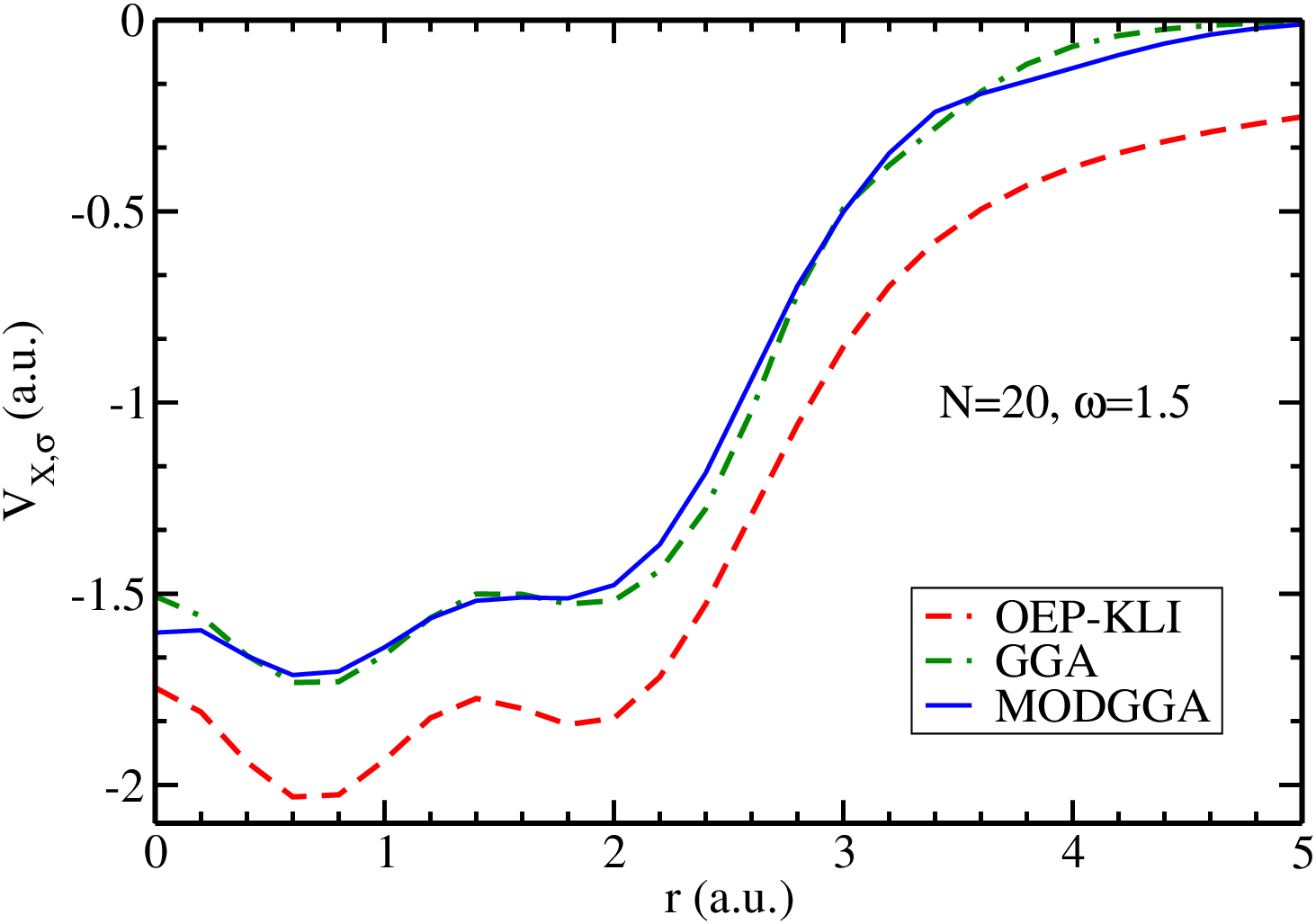} &
    \includegraphics[width=3.4in,height=2.0in,angle=0.0]{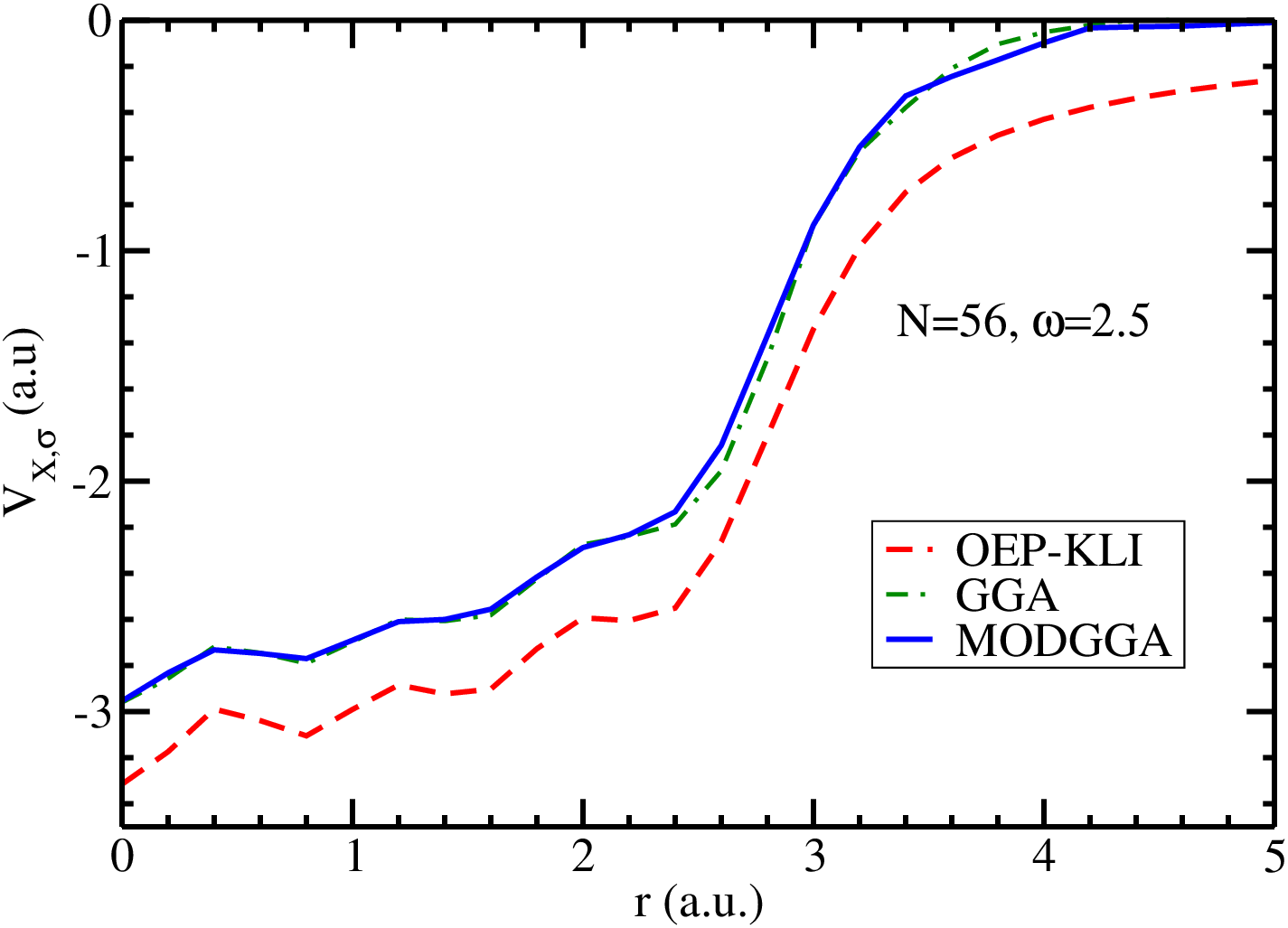}\\   
    \includegraphics[width=3.4in,height=2.0in,angle=0.0]{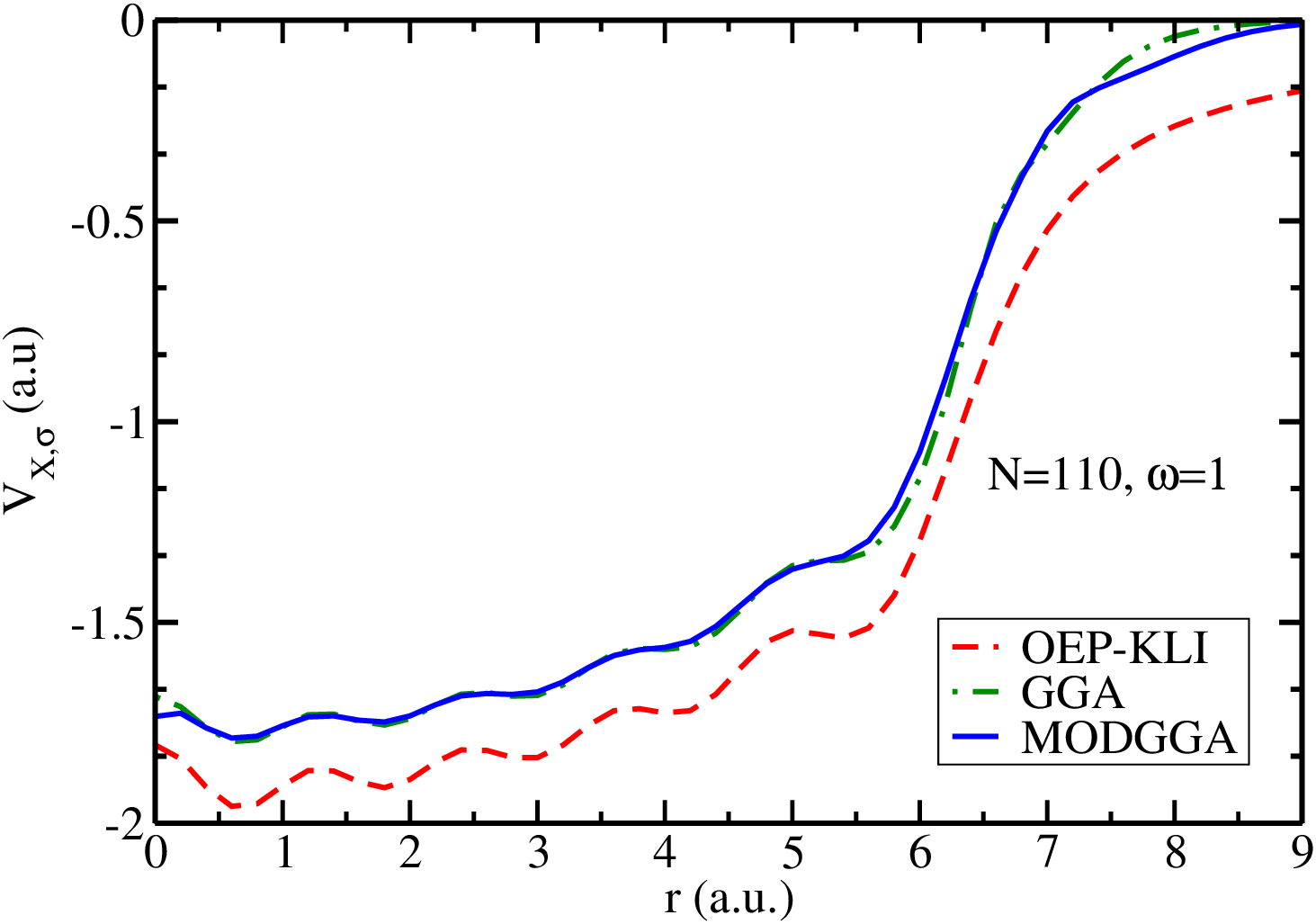} &
    \includegraphics[width=3.4in,height=2.0in,angle=0.0]{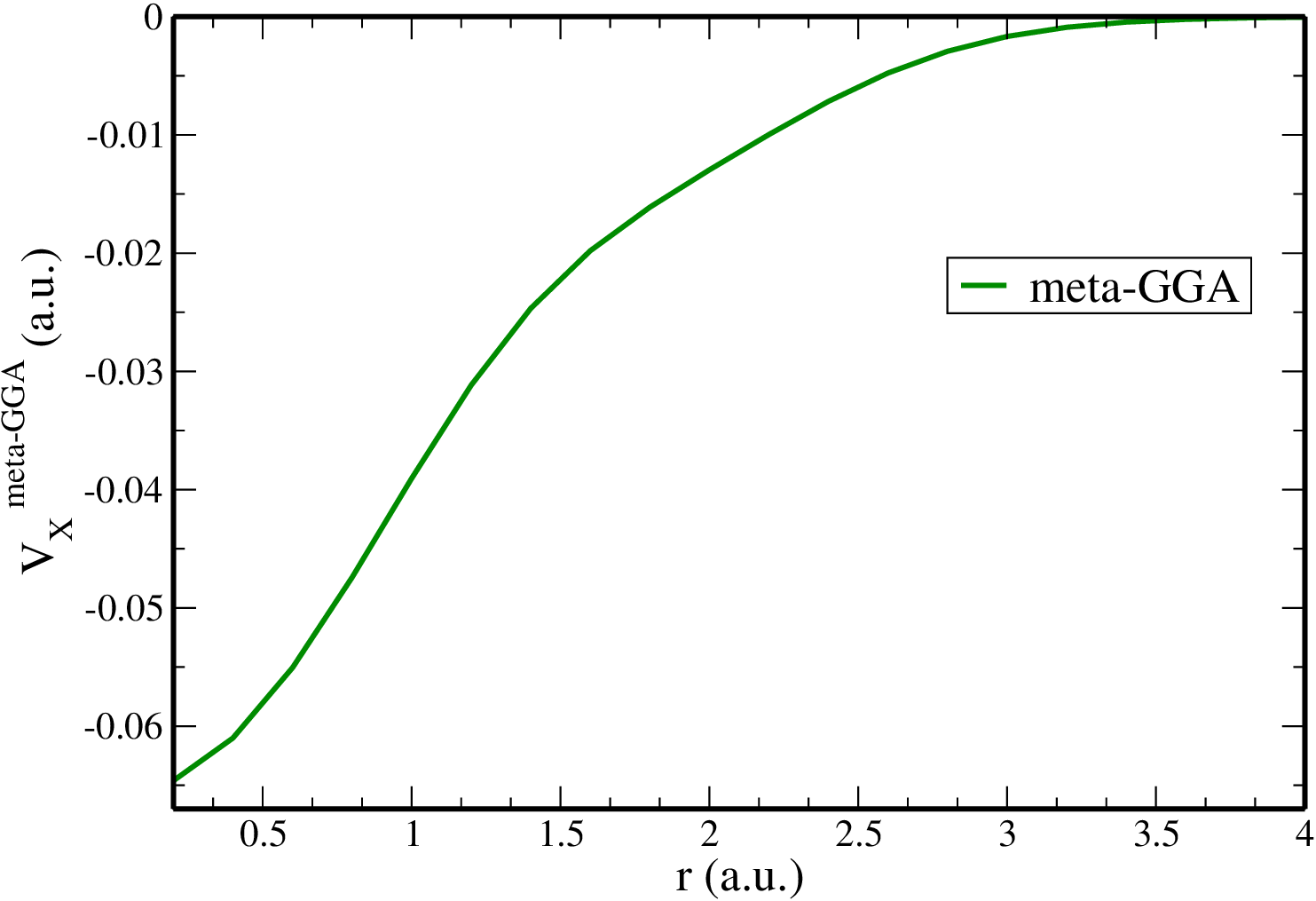} 
  \end{tabular}
  \caption{First five figures represent exchange potentials $v_{x\sigma}^{MOD-GGA}$ of Eq.(\ref{ap2eq3}) 
compared with GGA potential \cite{prvm} and KLI-OEP for different number of electrons and confinement strength 
$\omega$ in parabolic potential. Last figure represents potential of Eq.(\ref{ap3eq7}) for two non-interacting 
electron confined in a parabolic potential.}
\end{figure*}

\section{\label{sec:con}Concluding Remarks}
To summarize, we have obtained two forms of the enhancement factor and therefore the corresponding semilocal exchange 
energy functionals using ingredient of $GGA$ and inhomogeneity parameter defined by Becke. The newly constructed 
functionals have reduced the errors for the overall test set i.e., parabolic quantum dot with varying particle number 
and confinement strength compared to existing ones i.e., 2D-GGA, 2D-B88, 2D-BR. We have also obtained the form of exchange 
potential for our proposed functionals. The parameters of the functionals are obtained by using low and high density limits 
of the enhancement factor and by fitting with the exact exchange results of the parabolic quantum dot. So the proposed 
functionals will enable us for precise many-electron calculations of larger structures such as arrays of quantum dots and 
quantum-Hall devices. We believe that, the construction takes the GGA proposed by R\"{a}s\"{a}nan et. al one step forward 
in view of the improvement in the exchange energy. We have also shown that using inhomogeneity parameter of Becke, semilocal 
density functionals with improve performance can be constructed. In this way, one can propose hybrid density functional for 
2D systems as it has already been designed accurately in 3D.    

\section{\label{sec:ack}Acknowledgments}
The authors would like to acknowledge for the financial support from the Department of Atomic Energy, Government of India

\appendix
\section{\label{sec:factor}Coordinate Transformed Exchange-Hole Based Enhancement factor}\label{ap1}
Since the exchange energy is related to exchange hole and exchange hole is related first order reduced density 
matrix. Thus, the different form of exchange hole density plays a significant role in designing the exchange
energy functional. It is of great interest to study the short-range behavior of exchange hole under general 
coordinate transformation i.e., $(\vec{r}_1,\vec{r}_2)\to(\vec{r}^\lambda,u)$, where, $\vec{r}^\lambda = 
\lambda\vec{r}_1+(1-\lambda)\vec{r}_2$. Now the exchange energy functional becomes
\begin{equation}
E^{2D}_x=\frac{1}{2}\int~d^2r^\lambda\rho(\vec{r}^\lambda)\int~d^2u\frac{\rho^t_{x2D}(\vec{r}^\lambda,u)}
{u}
\label{ap1eq1}
\end{equation}
where, $\rho^t_{x2D}$ is the transformed exchange hole density defined by
\begin{equation}
\rho^t_{x2D}=-\frac{|\varGamma_{1t}^{2D}(\vec{r}^\lambda-(1-\lambda)\vec{u},\vec{r}^{\lambda}+\lambda
\vec{u})|^2}{2\rho(\vec{r})}
\label{ap1eq2}
\end{equation}
with $\varGamma_{1t}^{2D}$, be the KS single particle density matrix. The real parameter $\lambda$ takes
the value $1$ for conventional exchange hole and $\frac{1}{2}$ for on the top of the exchange hole. 
Now the transformed single particle KS density matrix is expressed 
around $u=0$ as
\begin{equation}
\begin{split}
\varGamma_{1t}^{2D}(\vec{r},\vec{u})=e^{\vec{u}.[-(1-\lambda)\vec{\nabla}_1+\lambda\vec{\nabla}_2]}
\varGamma_{1t}^{2D}(\vec{r},\vec{u})|_{\vec{u}=0}\\
=e^{\vec{u}.[-(1-\lambda)\vec{\nabla}_1+\lambda\vec{\nabla}_2]}
\sum_i^{occ}\Psi^{*}_i(\vec{r}^\lambda
-(1-\lambda)\vec{u})\\ \Psi_i(\vec{r}^{\lambda}+\lambda\vec{u})|_{\vec{u}=0}
\label{ap1eq3}
\end{split}
\end{equation}
where, $\vec{\nabla}_1$ and $\vec{\nabla}_2$ operate on $\Psi^{*}_i$ and $\Psi_i$ respectively. Taking the 
cylindrical average of Taylor series expansion of Eq.(\ref{ap1eq3}) yields the correct small $u$ behavior 
i.e.
\begin{equation}
\begin{split}
\langle\rho^t_{x2D}\rangle = -\frac{\rho(\vec{r})}{2}-\frac{1}{4}\Big[\Big(\lambda^2-\lambda+\frac{1}{2}
\Big)\nabla^2\rho(\vec{r})-2\tau\\
+\frac{1}{4}\Big(2\lambda-1\Big)^2\frac{|\vec{\nabla}\rho(\vec{r})|^2}
{\rho(\vec{r})}\Big]u^2
\end{split}
\label{ap1eq4}
\end{equation} 
Define a dimensionless parameter $Q_B^\lambda$,
\begin{equation}
\begin{split}
Q_B^\lambda = \frac{1}{\tau_0}\Big[\frac{1}{2}\Big(\lambda^2-\lambda+\frac{1}{2}
\Big)\nabla^2\rho+\tau_0\\
-\tau+\frac{1}{8}(2\lambda-1)^2\frac{|\vec{\nabla}\rho|^2}{\rho}\Big]
\end{split}
\label{ap1eq5}
\end{equation}
Now using Eq.(\ref{ap1eq5}), Eq.(\ref{ap1eq4}) can be written as,
\begin{equation}
\langle\rho^t_{x2D}\rangle = -\frac{\rho(\vec{r})}{2}+\frac{1}{2}\tau_0\Big(1-Q_B^\lambda\Big)u^2
\label{ap1eq6}
\end{equation}
Thus instead of Becke’s inhomogeneity parameter $Q_B$ a $\lambda$-dependent inhomogeneity parameter can be used in 
Eq.(\ref{eq19}) that leads to a family of enhancement factors.

\section{\label{sec:potential}Ingredients of GGA Potential}\label{ap2}
Here we have derived an explicit expression for the modified exchange potential, $v_x^{MOD-GGA}$ . As from exact spin scaling, 
the spin-labeled exchange potential is given by
\begin{equation}
\begin{split}
v_{x\sigma}^{2D-GGA}=&\frac{\delta E_x^{2D-GGA}[\rho_\uparrow,\rho_\downarrow]}{\delta\rho_\sigma(\vec{r})}\\
=&\frac{\delta E_x^{2D-GGA}[\rho_\uparrow,\rho_\downarrow]}{\delta\rho_\sigma(\vec{r})}\Big|_{\rho(\vec{r})
=2\rho_\sigma(\vec{r})}~,
\end{split}
\label{ap2eq1}
\end{equation}
where $\sigma=\uparrow$ or $\downarrow$ and $\rho(\vec{r})=\rho_\uparrow+\rho_\downarrow$. With MOD-GGA enhancement factor 
the exchange potential becomes
\begin{equation}
\begin{split}
v_{x\sigma}^{MOD-GGA} = A_x2^{1/2}\rho_{\sigma}(\vec{r})^{1/2}\Big[\frac{3}{2}F_x(s_{\sigma})\Big] + A_x2^{1/2}
\rho_{\sigma}(\vec{r})^{1/2}\\
\Big[-\frac{3}{2}s_{\sigma}(\vec{r})-\frac{1}{2k_F}\frac{\nabla^2\rho}{|\nabla\rho|}+\frac{1}{2k_F}\frac{\vec{\nabla}
\rho_\sigma\cdot\vec{\nabla}|\vec{\nabla}\rho_\sigma|}{|\vec{\nabla}\rho_\sigma|^2}\Big]\frac{dF{_x(s_\sigma)}}{ds_\sigma}\\
+A_x2^{1/2}\rho_{\sigma}(\vec{r})^{1/2}\Big[-\frac{1}{(2k_F)^2}\frac{\vec{\nabla}\rho\cdot\vec{\nabla}|\vec{\nabla}\rho|}
{|\vec{\nabla}\rho|\rho}+\frac{3}{2}s_{\sigma}^2\Big]\frac{d^2F_x(s_{\sigma})}{ds_{\sigma}^2}~,
\end{split}
\label{ap2eq3}
\end{equation}
where $A_x = \frac{4(2\pi)^{\frac{1}{2}}}{3\pi}$ and the enhancement factor and it's derivatives w.r.t reduced density 
gradient are given by
\begin{equation}
F_x(s_\sigma)=1+\mu\frac{s_\sigma\log(g_{\sigma})}{1+\beta s_{\sigma}^{\frac{1}{2}}\log(g_\sigma)+(1-e^{-c s_{\sigma}^2})}
\label{ap2eq4}
\end{equation}
\begin{equation}
\begin{split}
\frac{dF{_x(s_\sigma)}}{ds_\sigma}=\mu\frac{\log(g_\sigma)+\frac{s_\sigma}{g_\sigma}\frac{dg_{\sigma}}{ds_\sigma}}
{1+\beta s_{\sigma}^{\frac{1}{2}}\log(g_\sigma)+(1-e^{-c s_{\sigma}^2})}\\
-\mu\frac{s_{\sigma}\log(g_\sigma)[\frac{\beta}{2}s_{\sigma}^{-1/2}\log(g_\sigma)+\beta s_{\sigma}^{1/2}\frac{1}{g_\sigma}
\frac{dg_{\sigma}}{ds_\sigma}+2cs_\sigma e^{-cs_\sigma^2}]}{[1+\beta s_{\sigma}^{\frac{1}{2}}\log(g_\sigma)+
(1-e^{-c s_{\sigma}^2})]^2}
\end{split}
\label{ap2eq5}
\end{equation}
and
\begin{widetext}
\begin{eqnarray}
\frac{d^2F_x(s_{\sigma})}{ds_{\sigma}^2}&=&\mu\frac{\frac{2}{g_\sigma}\frac{dg_{\sigma}}{ds_\sigma}+\frac{s_\sigma}
{g_\sigma^2}\Big(\frac{dg_\sigma}{ds_\sigma}\Big)^2+\frac{s_\sigma}{g_\sigma}\frac{d^2s_\sigma}{ds^2_\sigma}}
{1+\beta s_{\sigma}^{\frac{1}{2}}\log(g_\sigma)+(1-e^{-c s_{\sigma}^2})}
-2\mu\frac{\Big[\log(g_\sigma)+\frac{s_\sigma}{g_\sigma}\frac{dg_{\sigma}}{ds_\sigma}\Big][\frac{\beta}{2}s_{\sigma}
^{-1/2}\log(g_\sigma)
+\beta s_{\sigma}^{1/2}\frac{1}{g_\sigma}\frac{dg_{\sigma}}{ds_\sigma}+2cs_\sigma e^{-cs_\sigma^2}]}
{[1+\beta s_{\sigma}^{\frac{1}{2}}\log(g_\sigma)+(1-e^{-c s_{\sigma}^2})]^2}\nonumber\\
&-&\mu \frac{s_\sigma\log(g_\sigma)[-\frac{\beta}{4}s_\sigma^{-3/2}\log(g_\sigma)
+\beta s_\sigma^{-1/2}\frac{1}{g_\sigma}\frac{dg_\sigma}{ds_\sigma}
-\beta s_\sigma^{1/2}\frac{1}{g_\sigma^2}(\frac{dg_\sigma}{ds_\sigma})^2
+\frac{\beta}{g_\sigma}\frac{d^2g_\sigma}{ds_\sigma^2}
-2ce^{-cs_\sigma^2}-4c^2 s_\sigma^2e^{-cs_\sigma^2}]}
{[1+\beta s_{\sigma}^{\frac{1}{2}}\log(g_\sigma)+(1-e^{-c s_{\sigma}^2})]^2}\nonumber\\
&+&2\mu\frac{s_{\sigma}\log(g_\sigma)[\frac{\beta}{2}s_{\sigma}^{-1/2}\log(g_\sigma)+\beta s_{\sigma}^{1/2}\frac{1}
{g_\sigma}\frac{dg_{\sigma}}{ds_\sigma}+2cs_\sigma e^{-cs_\sigma^2}]^2}{[1+\beta s_{\sigma}^{\frac{1}{2}}\log(g_\sigma)
+(1-e^{-c s_{\sigma}^2})]^3}~.
\end{eqnarray}
\end{widetext}

\section{\label{sec:mgga}Ingredients of Two Electrons Meta-GGA Potential}\label{ap3}
If any general density functional is given by
\begin{equation}
F[\rho]=\int~d^2r~G[\rho,\vec{\nabla}\rho,\nabla^2\rho,.......,\nabla^m\rho;\vec{r}]~.
\label{ap3eq1}
\end{equation}
Then, the functional derivative of the above functional is
\begin{equation}
\begin{split}
\frac{\delta F[\rho]}{\delta\rho}=\frac{\partial G[\rho]}{\partial\rho}-\vec{\nabla}\rho\cdot\frac{\partial G}{\partial
\vec{\nabla}\rho}+\vec{\nabla}^2\rho\cdot\frac{\partial G}{\partial\vec{\nabla}^2\rho}+...\\
+(-1)^m\vec{\nabla}^m\rho\cdot\frac{\partial G}{\partial
\vec{\nabla}^m\rho}~.
\end{split}
\label{ap3eq2}
\end{equation}
So for the exchange energy functional
\begin{equation}
E_x[\rho] = {A_x} \int~d^2r~\rho(\vec{r})^{\frac{3}{2}}F_x[Q_B]~,
\label{ap3eq3}
\end{equation}
the corresponding exchange potential is given by 
\begin{equation}
\begin{split}
\frac{v_x}{A_x}=\frac{3}{2}\rho(\vec{r})^{\frac{1}{2}}F_x[Q_B]+\rho(\vec{r})^{\frac{3}{2}}\frac{dF_x}{dQ_B}\frac{\partial Q_B}
{\partial\rho}\\
-\vec{\nabla}\cdot\Big[\rho(\vec{r})^{\frac{3}{2}}\frac{dF_x}{dQ_B}\frac{\partial Q_B}{\partial\vec{\nabla}\rho}\Big]
+\vec{\nabla}^2\cdot\Big[\rho(\vec{r})^{\frac{3}{2}}\frac{dF_x}{dQ_B}\frac{\partial Q_B}{\partial\vec{\nabla}^2\rho}\Big]
\end{split}~.
\label{ap3eq4}
\end{equation}
As for two electron systems,
\begin{equation}
\tau=\frac{1}{8}\frac{|\vec{\nabla}\rho|^2}{\rho}~.
\label{ap3eq5}
\end{equation} 
Therefore,
\begin{equation}
Q_B = 1 + \frac{1}{4\tau_0}\nabla^2\rho = 1 + \frac{1}{4c_f}\frac{\nabla^2\rho}{\rho^{\frac{3}{2}}}~,
\label{ap3eq6}
\end{equation}
where $c_f=\frac{1}{4}(2\pi)^{\frac{1}{2}}$. Now using Eq.(\ref{ap3eq6}) into Eq.(\ref{ap3eq4}) exchange only potential 
is obtained to be
\begin{equation}
v_x=A_x\Big[\frac{3}{2}\rho(\vec{r})^{\frac{1}{2}}F_x-\frac{3}{8c_f}\frac{\nabla^2\rho}{\rho}\frac{dF_x}{dQ_B}
+\frac{1}{4c_f}\nabla^2\Big[\frac{dF_x}{dQ_B}\Big]\Big]
\label{ap3eq7}
\end{equation}
with
\begin{equation}
\frac{dF_x}{dQ_B}=\frac{\alpha}{\sqrt{1+(\gamma Q_B)^2}}-\frac{\alpha\gamma^2 Q_B^2}{[1+(\gamma Q_B)^2]^{3/2}}~.
\label{ap3eq8}
\end{equation}
For the case of two non-interacting electrons confined in a parabolic quantum dot, the electron density is given by
\begin{equation}
\rho(\vec{r}) = \frac{2}{\pi}\exp(-2r^2) 
\label{ap3eq9}
\end{equation}
In cylindrical coordinate the Laplacian operator $\nabla^2$ can be written as
\begin{equation}
\nabla^2 = \frac{\partial^2}{\partial r^2} + \frac{1}{r}\frac{\partial}{\partial r}
\label{ap3eq10}
\end{equation}
So by making use of the above density and the Laplacian in $Q_B$ one can obtain $v_x$ for a non-interacting two-electron system.

\end{document}